\documentstyle[12pt]{article}

\renewcommand{\thesection}{\Roman{section}.}
\renewcommand{\theequation}{\Roman{section}.\arabic{equation}}
\makeatletter
\@addtoreset{equation}{section}
\makeatother

\newcommand{\Ga}{\Gamma}

\newcommand{\al}{\alpha}
\newcommand{\be}{\beta}
\newcommand{\la}{\lambda}

\newcommand{\de}{\delta}
\newcommand{\pa}{\partial}

\newcommand{\si}{\sigma}

\newcommand{\varep}{\varepsilon}

\newcommand{\sq}{\sqrt}
\newcommand{\ov}{\over}
\newcommand{\ab}{{\alpha\beta}}
\newcommand{\ba}{{\beta\alpha}}

\newcommand{\oot}{{1\ov 2}}
\newcommand{\oof}{{1\ov 4}}
\newcommand{\ooe}{{1\ov 8}}
\newcommand{\mn}{{\mu\nu}}

\newcommand{\pp}[2]{{\partial #1\over\partial #2}}

\newcommand{\eo}{E_{1 \ \si}^{\ \mu}} 
\newcommand{\ez}{E_{0 \ \si}^{\ \mu}} 

\begin{document}

\title{The Energy-Momentum Tensor for Gravitational Interactions}

\author{Walter Wyss \\ \\
        Department of Physics \\ 
        University of Colorado \\ 
        Boulder, CO 80309}

\date{}

\maketitle

\begin{abstract}
Within the Lagrange formalism we show that the gauge invariant total 
energy-momentum tensor for gravitational interactions is zero. 
If the equations of motion are satisfied the energy tensor is conserved. 
\end{abstract}

\clearpage

\section{Introduction}
 
All our considerations are within the Lagrange formalism of field theory. 
The concept of the energy-momentum tensor for gravitational interactions 
has a long history.  In this paper we adopt the philosophy commonly 
accepted for electromagnetic interactions.  The electromagnetic field 
is described by a restmass zero, spin one field.  This field then 
has its own gauge group.  A charged matter field also has its own 
gauge group, that leads to a conserved current.  If we demand that 
the electromagnetic interaction with the charged matterfield is a 
minimal coupling to the conserved current, the two gauge 
groups get related to form the EM interaction gauge group 
(electromagnetic gauge group).  It consists of scalar functions and 
gives rise to a covariant derivative.  The total action is then 
Lorentz-invariant and also EM gauge invariant.  Similarly, the  
gravitational field is described by a restmass zero, spin two 
field.  This field also has its own gauge group.  If we demand that  
the gravitational interaction with a matterfield is a minimal coupling 
to the conserved energy-momentum tensor of the matterfield, the gauge 
group gets enlarged to the $G$ interaction gauge group.  Its 
Lie algebra consists of vector fields and gives raise to a covariant 
derivative.  The total action is then Lorentz invariant and 
also $G$-gauge invariant [2] (gravitational gauge group). 
 
We take the position that the Lorentz group is the fundamental symmetry 
group of all of physics.  If there are additional symmetry groups, the 
physical relevant quantities should be covariant with respect to the 
additional symmetry.  Conservation laws can involve quantities that 
are not covariant with respect to the additional symmetry groups 
but are only Lorentz covariant. 
 
In this paper we study gravitational interactions with a general matter 
field.  The $G$ interaction gauge group consists of vector fields 
vanishing at infinity.  The Lorentz group is thus not a subgroup of 
our $G$ gauge group.  With respect to the Lorentz group, the energy 
tensor (translation invariance) and the energy-momentum tensor 
(proper Lorentz invariance) are of physical importance [3].  We show 
that the energy tensor satisfies a conservation law and the $G$ 
gauge invariant energy-momentum tensor vanishes.  $G$  gauge invariance 
implies additional conservation laws. 
 
\section{Gravitational Interaction with a General Matterfield} 
 
Let $g_\ab$ be a symmetric Lorentz tensor field on Minkowski space and 
$g^\ab$ its inverse.  $\phi$ stands for a general multicomponent  
field on Minkowski space.  We assume throughout automatic summation 
over Lorentz-, spinor-, and internal indices even if they are not 
explicitly mentioned.  The action for gravitational interaction is 
given by 
\begin{equation} 
A = \int dx L 
\end{equation} 
where 
\begin{equation} 
L = L_G(g) + L_M (g, \phi). 
\end{equation} 
Let 
\begin{equation} 
A_G = \int dx L_G (g) 
\end{equation} 
and 
\begin{equation} 
A_{GM} = \int dx L_M (g, \phi) 
\end{equation} 
Both actions $A_G$ and $A_{GM}$ are assumed to be Lorentz invariant. 
 
In order for the total action $A$ to represent gravitational 
interaction [2], the action also has to be invariant with respect 
to the $G$ gauge group (gravitational gauge group).  Its infinitesimal 
generators (Lie algebra) consist of smooth vector fields, vanishing 
at infinity.  This invariance is represented by smooth coordinate 
transformations, i.e., general covariance. 
 
The action $A_G$ represents gravitational selfinteraction.  Its 
Lorentz invariance gives an energy tensor (translation invariant) and 
an energy-momentum tensor.  The same is true for the action $A_{GM}$ 
which represents the gravitational interaction with matter.  The actions 
$A_G$ and $A_{GM}$ are in additional also $G$-gauge invariant.

In the next two sections we compute the energy tensors and 
energy-momentum tensors for these actions and the 
implications of the additional $G$-gauge symmetry. 
 
\section{Gravitational Selfinteraction} 
 
With the notations and the results in the appendices and form [3] and 
[4] the action for gravitational selfinteraction is given by 
\begin{equation} 
A_G = \int dx L_G (G) 
\end{equation} 
where the Lagrangian is given by 
\begin{equation} 
L_G (G) = \sq{g} K + \pa_\mu B^\mu. 
\end{equation} 
The divergence term is resopnsible for the action to be $G$-gauge 
invariant.  Deleting the divergence results in a Lorentz invariant 
action only. 
 
We now find the following quantities 
\begin{equation} 
H_0^{\ \ab,\mu} \equiv 
\pp {(\sq{g}K)} {g_\ab,\mu} 
\end{equation} 
\begin{equation} 
H_0^{\ \ab,\mu} = - \oot g^\ab B^\mu 
+ \sq{g} 
\left[ \Ga^{\mu\ab} - \oot  
\left(g^{\mu\al} \Ga^{\si\be}_{\ \ \si} + g^{\mu\be} \Ga^{\si\al}_{\ \ \si} 
\right) \right]. 
\end{equation} 
The Euler derivative is given by 
\begin{eqnarray} 
G_0^{\ \ab} & \equiv & \varep (g_\ab) \sq{g} K \\
G_0^{\ \ab} &     =  & - \sq{g} \left[ R^\ab - \oot g^\ab R \right] 
\end{eqnarray}
 
The quantity in the bracket is also known as the Einstein tensor. 
 
The energy tensor is given by  
\begin{equation} 
E_{0 \ \si}^{\ \mu} = H_o^{\ \ab,\mu} 
g_{\ab,\si} - \de^\mu_{\ \si} \sq{g} K 
\end{equation} 
Introducing 
\begin{equation} 
h^\ab = \sq{g} g^\ab\ , \ h_\ab =  
{1 \ov \sq{g}} g_\ab 
\end{equation} 
we get 
\begin{equation} 
E_{0\ \si}^{\ \mu} = \Ga^\be_{\ \ab} 
\pa_\si h^\ab - \Ga^\mu_{\ \ab} 
\pa_\si h^\ab - \de^\mu_{\ \si} \sq{g} K 
\end{equation} 
$E_{0 \ \si}^{\ \mu}$ is not $G$-gauge invariant and thus is only a  
Lorentz tensor. 
 
From translation invariance we find 
\begin{equation} 
\pa_\mu E_{0 \ \si}^{\ \mu} = - G^\ab_0 g_{\ab,\si} 
\end{equation} 
Now we look at the other auxiliary quantities and find 
\begin{eqnarray} 
P_{0 \ \ \, \la}^{\ \mu\al}  &=& 0 \\
K_{0 \ \ \, \la}^{\ \mu\al}  &=& 2H_0^{\ \ab,\mu} g_{\be\la} \\
Z_{0 \ \si}^{\ \mu} &= &0 
\end{eqnarray}
From $G$-gauge invariance one finds 
\begin{equation} 
\oot \left[  
K_{0 \ \ \la} ^{\ \ab} + 
K_{0 \ \ \la} ^{\ \ba}  \right] 
= - \pa_\mu 
\left[ \pp {B^\mu} {g_{\al\varrho,\be}} g_{\varrho \la} 
+  
\pp {B^\mu} {g_{\be\varrho,\al}} g_{\varrho \la} \right] 
\end{equation} 
and from 
\begin{equation} 
\pp {B^\mu} {g_{\al\varrho,\be}} g_{\varrho \la}  
= - \de^\al_{\ \la} h^{\mu\be} + \oot 
\de^\be_{\ \la} h^{\mu\al} + \oot g^\mu_{\ \la} 
h^\ab 
\end{equation} 
we then get 
\begin{equation} 
\oot \left[  
K_{0 \ \ \la} ^{\ \ab} + 
K_{0 \ \ \la} ^{\ \ba}  \right] 
=  \pa_\mu 
\left[  
\oot \de^\al_{\ \la} h^{\mu\be} 
+ \oot \de^\be_{\ \la} h^{\mu\al} 
- \de^\mu_{\ \la} h^{\ab} \right]. 
\end{equation} 
Observe now that 
\begin{equation} 
\pa_\al \pa_\be K_{0\ \ \, \la}^{\ \ab} = 0 
\end{equation} 
Also 
\begin{equation} 
W_0^{\ \la\mu\al} = \pa_\si 
\left[ 
\eta^{\al\si} h^{\mu\la} + \eta^{\mu\si} h^{\al\la} 
-\eta^{\mu\al} h^{\mu\la} - \eta^{\si\la} h^{\mu\al} 
\right] 
\end{equation} 
gives 
\begin{equation} 
t^{\mu \al} 
= \pa_\la \pa_\si 
\left[ 
\eta^{\al\si} h^{\mu\la} + \eta^{\mu\si} h^{\al\la} 
-\eta^{\mu\al} h^{\si\la} - \eta^{\si\la} h^{\mu\al} 
\right] 
\end{equation} 
Since $\pa_\al \pa_\be K_{0 \ \ \, \la}^{\ \ab} = 0$ and 
$t^{\mu\al}$ is only a Lorentz tensor but not $G$-gauge invariant, 
the $G$-gauge invariant energy-momentum tensor is identically zero. 
 
This also implies that a $G$-gauge invariant action, whose Lagrangian 
involves only up to first order derivatives of the fields and no 
divergence term, has energy-momentum tensor zero. 
 
That the energy-momentum tensor vanishes identically is expressed by 
the identity 
\begin{equation} 
E_{0 \ \si}^{\mu} + 2 G_{0}^{\ \mu\al} g_{\al\si} + 
\pa_\la K_{0 \ \ \, \si}^{\ \la\mu} \equiv 0 
\end{equation} 
Finally we get from $G$ gauge invariance the identity 
\begin{equation} 
2 \pa_\be \left(G_0^{\ \ba} g_{\al\la} \right)  
= G_0^{\ \ab} g_{\ab,\la} 
\end{equation} 
This is the Bianchi identity. 
 
\section{Gravitational Interaction Term with Matter} 
 
The gravitational interaction term with matter is given by 
the action  
\begin{displaymath} 
A_{GM} = \int dx L_M (g, \phi) 
\end{displaymath} 
This action is Lorentz invariant and also $G$-gauge invariant. 
 
The Lagrangian  
\begin{equation} 
L_M (g,\phi) = L_M (g_\ab, \pa_\mu g_\ab, \phi, \pa_\si \phi) 
\end{equation} 
is assumed to depend only on the fields $g_\ab, \phi$ and on their 
first derivatives, and has no boundary term.  Thus the energy-momentum 
tensor vanishes identically. 
 
We now find the following quantities 
\begin{equation} 
H_M^{\ \  \ab, \mu} \equiv 
\pp {L_M} {g_{\ab,\mu}} 
\end{equation} 
This can only be evaluated if the matter fields $\phi$ are specified. 
 
Similarly for 
\begin{equation} 
H_M^{\ \  \mu} \equiv \pp {L_M} {\phi_\mu} 
\end{equation} 
The Euler derivatives are given by 
\begin{equation} 
\varep (g_\ab) L_M \equiv M^\ab 
\end{equation} 
We call $M^\ab$ the gravitational stress tensor. 
\begin{equation} 
G \equiv \varep (\phi) L_M. 
\end{equation} 
is the Euler derivative of the matter fields.  The energy tensor for 
the action $A_{GM}$ is given by 
\begin{equation} 
\eo = E_{M \ \, \ \si}^{\ \ \, \mu}  
+ H^{\ \ \ab,\mu}_M  g_{\ab, \si} 
\end{equation} 
where 
\begin{equation} 
E_{M \ \si}^{\ \mu} =  
H_{M \ \, \si}^{\ \  \mu}   \phi_\si  
- \de^\mu_{\ \si} L_M 
\end{equation} 
is the energy tensor for the matter field.  Again, both $ 
E_{1 \ \si}^{\ \mu}$ and $ E_{M \ \, \si}^{\ \  \mu}$ are only 
Lorentz tensors, because in general they are not $G$-gauge invariant.

From translation invariance we find 
\begin{equation} 
\pa_\mu \eo = - G \phi_\si - M^\ab g_{\ab,\si} 
\end{equation} 
Now we look at the other auxiliary quantities and find 
\begin{equation} 
K_{1 \ \ \, \la}^{\ \mu\al} = H^\mu 
S^\al_{\ \la} + 2 H_M^{\ \  \ab,\mu} g_{\be\la} 
\end{equation} 
and 
\begin{equation} 
Z_{1 \ \si} ^{\ \mu} = \eo + GS^\mu_{\ \si}  
+ 2M^{\mu\be} g_{\be\si} + \pa_\la  
K_{1 \ \ \si}^{\ \la\mu} 
\end{equation} 
Since $Z_{1\ \si}^{\ \mu}$ vanishes identically we obtain 
\begin{equation} 
E_{M\ \,\si}^{\ \  \mu} + GS^\mu_{\ \si} 
+ \pa_\la (H^\la S^\mu_{\ \si}) 
+ 2 M^{\mu\be} g_{\be\si} + H_M^{\ \  \ab,\mu} 
g_{\ab, \si} + 2 \pa_\la 
\left[ H_M^{\ \  \mu\be,\la} g_{\be\si} \right] = 0 
\end{equation} 
From  $G$-gauge invariance we finally get the identity 
\begin{equation} 
\pa_\be \left[ GS^\be_{\ \la} + 2 M^{\ba} g_{\al\la} \right] 
= G \phi_\la + M^\ab g_{\ab,\la} 
\end{equation} 
 
\section{General Gravitational Interaction} 
 
We now look at the total action  
\begin{equation} 
A = A_G + A_{GM} 
\end{equation} 
with the corresponding Lagrangian 
\begin{equation} 
L = L_G + L_M 
\end{equation} 
The auxiliary quantities then read 
\begin{eqnarray} 
H^{\ab,\mu} & \equiv & \pp {[\sq{g} K + L_M]} {g_{\ab,\mu}} \\
H^{\ab,\mu} &=& H_0^{\ \ab,\mu} + H_M^{\ \  \ab,\mu} \\
H^\mu &\equiv& \pp {L_M} {\phi_\mu} \\
H^\mu &=& H_M^{\ \  \mu} 
\end{eqnarray}
For the Euler derivatives we get 
\begin{eqnarray} 
G^\ab &\equiv& \varep (g_\ab) [\sq{g} K + L_M] \\
G^\ab &=& G_0^{\ \ab} + M^\ab \\
G &\equiv& \varep (\phi) L_M 
\end{eqnarray}
The total energy tensor then becomes 
\begin{equation} 
E^\mu_{\ \si} = \ez + E_{M \ \si}^{\ \  \mu}  
+ H_M^{\ \  \ab,\mu} g_{\ab,\si} 
\end{equation} 
where $\ez$ is given by (III.9) and $E_{M\ \si}^{\ \  \mu}$ by (IV.7). 
 
$E^\mu_{\ \si}$ is only a Lorentz tensor. 
 
From translation invariance we get 
\begin{equation} 
\pa_\mu E^\mu_{\ \si} = - G^\ab g_{\ab,\si} - G \phi_\si 
\end{equation} 
The $G$-gauge invariant energy-momentum tensors belonging to the  
actions $A_G$ and $A_{GM}$ are both identically zero.  Thus the 
overall $G$-gauge invariant energy-momentum tensor is zero. 
 
This is reflected in the following two identities 
\begin{equation} 
\ez + 2G_0^{\ \mu\al} g_{\al\si} \pa_\la K_{0 \ \ \si}^{\ \la\mu} = 0 
\end{equation} 
\begin{equation} 
E_{M \ \,\si}^{\ \,\mu} + GS^\mu_{\ \si} 
+ \pa_\la (H^\la S^\mu_{\ \si}) + 2M^{\mu\be} 
g_{\be\si} + H_M^{\ \  \ab,\mu} 
g_{\ab,\si} + 2 \pa_\la 
\left[ H_M^{\ \  \mu\be,\la} g_{\be\si} \right] =0 
\end{equation} 
Finally $G$-gauge invariance gives the two identities 
\begin{eqnarray} 
2 \pa_\be \left( G_0^{\ \ba} g_{\al\la} \right) 
&=& G_0^{\ \ab} g_{\ab,\la} \\ 
\pa_\be \left[ GS^\be_{\ \la} + 2 M^{\ba} g_{\al\la} \right] 
&=& G \phi_\la + M^\ab g_{\ab,\la}  
\end{eqnarray} 
Now we assume the equations of motion to be satisfied, i.e., 
\begin{eqnarray} 
G &=&0 \\
G^\ab &=& 0 
\end{eqnarray} 
$G = 0 $ is the equation of motion for the matterfield and 
$G^\ab = 0$ is the Einstein equation 
\begin{equation} 
\sq{g} \left[ R^\ab - \oot g^\ab R \right] = M^\ab 
\end{equation} 
From translation invariance we get the conservation law 
\begin{equation} 
\pa_\mu E^\mu_{\ \si} = 0 
\end{equation} 
This means that the total energy is conserved.  This statement is 
reflected in the equations 
\begin{eqnarray} 
\pa_\mu \ez &=& M^\ab g_{\ab, \si} \\
\pa_\mu E_{M\ \si}^{\ \  \mu} 
&=& - M^\ab g_{\ab,\si} - \pa_\mu 
\left[H_M^{\ \  \ab,\mu} g_{\ab,\si} \right] 
\end{eqnarray} 
Finally we have the energy-balance equation 
\begin{equation} 
E^\mu_{\ \si} + \pa_\la 
\left[ K_{0 \ \ \si}^{\ \la\mu} + K_{1 \ \ \si}^{\ \la\mu} \right] = 0 
\end{equation} 
which is reflected in the equations 
\begin{equation} 
\ez - 2M^{\mu\be} g_{\al\si} + \pa_\la K_{0 \ \ \si}^{\ \la\mu} = 0 
\end{equation} 
\begin{equation} 
E_{M \ \si}^{ \ \  \mu} + 2 M^{\mu\be} g_{\be\si} 
+ H_M^{\ \  \ab,\mu} g_{\ab,\si} 
+ \pa_\la 
\left[H^\la S^\mu_{\ \si} + 2 H_M^{\ \  \mu\be,\la} 
g_{\be\si} \right] = 0 
\end{equation} 
and from $G$-gauge invariance the relation 
\begin{equation} 
2\pa_\be \left[ M^{\be\al} g_{\al\la} \right] = M^\ab g_{\ab,\la} 
\end{equation} 
 
\section{Example: Gravitational Interaction with a Massive Vectorfield} 
 
The gravitational interaction term for this model is given by 
the Lagrangian 
\begin{equation} 
L_M (g, \phi_\al) = \sq{g} 
[g^\ab g^\mn (D_\mu \phi_\al) (D_\nu \phi_\be) - m^2 
g^\ab \phi_\al\phi_\be ] 
\end{equation} 
with the covariant derivative 
\begin{equation} 
D_\mu \phi_\al = \pa_\mu \phi_\al - \Ga^\si_{\ \al\mu} \phi_\si 
\end{equation} 
for the vectorfield $\{ \phi_\al\}$. 
 
We first compute the auxiliary quantities 
\begin{eqnarray} 
H_M^{\ \  \al,\mu} &\equiv& \pp {L_M} {\phi_{\al,\mu}}  \\
H_M^{\ \  \ab,\mu} &\equiv& \pp {L_M} {g_{\ab,\mu}} \\
H_M^{\ \  \al,\mu} &=& 2 \sq{g} D^\mu \phi^\al  \\
H_M^{\ \  \ab,\mu} &=& - \oot \sq{g} 
\big[ \phi^\al (D^\be \phi^\mu + D^\mu \phi^\be) + 
\phi^\be (D^\al \phi^\mu + D^\mu \phi^\al) \nonumber \\
& & -  \phi^\mu (D^\al \phi^\be + D^\be \phi^\al ) \big] 
\end{eqnarray} 
We have raised the indices with $g^\ab$. 
 
For the Euler derivatives 
\begin{eqnarray} 
G^\al &\equiv& \varep (\phi_\al) L_M \\
M^\ab &\equiv& \varep (g_\ab) L_M  
\end{eqnarray} 
we get 
\begin{eqnarray} 
G^\al &=& - 2 \sq{g} [ D_\mu D^\mu \phi^\al + m^2 \phi^\al] \\
M^\ab &=& \oot g^\ab L_M - \sq{g}  
[(D^\mu \phi^\al) (D_\mu \phi^\be) + (D^\al \phi^\mu) 
(D^\be \phi_\mu) - m^2 \phi^\al \phi^\be ]  \\
&+& \oot \sq{g} D_\mu 
\left[ \phi^\al (D^\be \phi^\mu + D^\mu \phi^\be) +\phi^\be  (D^\al \phi^\mu 
+ D^\mu \phi^\al) - \phi^\mu (D^\al \phi^\be + D^\be \phi^\al) \right] \nonumber
\end{eqnarray} 
The energy tensor for the matterfield alone is given by 
\begin{equation} 
E_{M\ \,\si}^{\ \ \mu} = 2 \sq{g} 
(D^\mu \phi^\al) \phi_{\al,\si} - \de^\mu_{\ \si} L_M 
\end{equation} 
The total energy tensor for this gravitational interaction then 
becomes 
\begin{equation} 
E^\mu_{\ \si} = \ez  + E_{M\ \,\si}^{\ \  \mu} +  
H_M^{\ \  \ab,\mu} g_{\ab,\si} 
\end{equation} 
where $\ez$ is given by (III.9) as 
\begin{equation} 
\ez= \Ga^\be_{\ \ab} \pa_\si h^{\al\mu} 
- \Ga^\mu_{\ \ab} \pa_\si h^\ab - \de^\mu_{\ \si} \sq{g} K 
\end{equation} 
 
The equations of motion now read 
\begin{equation} 
D_\mu D^\mu \phi_\al + m^2 \phi_\al = 0 
\end{equation} 
\begin{eqnarray}
R^\ab - \oot g^\ab R & = &\oot g^\ab 
\left[ (D_\mu \phi_\si) (D^\mu \phi^\si) - m^2 \phi_\si \phi^\si \right]  
\nonumber \\
&&- (D^\mu \phi^\al) (D_\mu \phi^\be) - (D^\al \phi^\mu)  
(D^\be \phi_\mu)  
\nonumber \\
&&+ m^2 \phi^\al \phi^\be + \oot D_\mu 
\big[ \phi^\al (D^\be \phi^\mu + D^\mu \phi^\be) + 
\phi^\be (D^\al \phi^\mu + D^\mu \phi^\al) 
\nonumber \\
&&- \phi^\mu (D^\al \phi^\be + D^\be \phi^\al) \big] 
\end{eqnarray}
In addition we have the conservation law 
\begin{equation} 
\pa_\mu E^\mu_{\ \si} = 0 
\end{equation} 
where $E^\mu_{\ \si}$ is given by (VI.12). 
 
For the gravitational stress tensor we also have the equation 
\begin{equation} 
2 \pa_\be \left[ M^\ba g_{\al\la} \right]  
= M^\ab g_{\ab,\la} 
\end{equation} 
which is equivalent to the Bianchi identity. 
 
\section{Conclusions} 
 
We consider Lorentz-invariance as the fundamental symmetry of all 
of physics.  Within the framework of the Lagrange Formalism a 
Lorentz-invariant action gives raise to the Energy tensor 
(due to translation invariance) and to the Energy-momentum tensor 
(due to proper Lorentz-rotations).  If the equations of motion are  
satisfied both these tensors are conserved.  For gravitational 
interactions there is the additional symmetry of $G$-gauge 
transformations vanishing at infinity. If the action if  
$G$-gauge invariant, the $G$-gauge covariant energy-momentum 
tensor vanishes identically.  As we will see in a forthcoming 
paper this statement will give a precise meaning of the folklore 
statement ``The right hand side of Einstein's equations is given by 
the energy-momentum tensor of matter."  The energy tensor for the 
gravitational interaction is however not $G$-gauge covariant. 
But if the equations of motion are satisfied, the energy tensor 
is conserved. 
 
We hope that this paper puts to rest all the speculative statements 
about the energy-momentum tensor in General Relativity. 
 
\clearpage

\renewcommand{\thesection}{Appendix \Alph{section}:}
\renewcommand{\theequation}{\arabic{equation}}
\setcounter{section}{0}
\section{Notations}
 
Let $\eta_\mn$ denote the Lorentz metric tensor with signature 
$(+, -, -, -)$ and $\eta^\mn$ its inverse.  The Lorentz  
4-volume element is represented by $dx$.  We use the abbreviation 
$\pa_\mu = {\pa \ov \pa x^\mu}$ and for any field $\phi$ on 
Minkowski space $\pa_\mu \phi = \phi_\mu$.  Let $g_\ab$ be 
a symmetric Lorentz tensor field on Minkowski space and  
$g^\ab$ be its inverse, i.e., 
\begin{displaymath} 
g^{\al \mu} g_{\mu\be} = \de^\al_{\ \be} 
\end{displaymath} 
We introduce the following quantities 
\begin{eqnarray*}
g &=& - Det (g_\mn) \\
\Ga_{\mu\ab} &=& \oot [g_{\mu\al,\be} + g_{\mu\be,\al} - g_{\ab,\mu}]\\ 
\Ga^\mu_{\ \ab} &=& g^{\mu\si} \Ga_{\si\ab} \\
\Ga^{\mu\si}_{\ \ \, \be} &=& g^{\si\al} \Ga^\mu_{\ \ab}\\ 
\Ga^{\mu\si\varrho} &=& g^{\varrho\be} \Ga^{\mu\si}_{\ \ \be} \\ 
K_\mn &=& \Ga^\be_{\ \mu\al} \Ga^\al_{\ \nu\be} - \Ga^\al_{\ \mn} \Ga^\be_{\ 
\ab}\\ 
B^\mu &=& \sq{g} [\Ga^{\mu\al}_{\ \ \al} - \Ga^{\al\mu}_{\ \ \al}] \\
R_{\mn} &=& \Ga^\al_{\ \mn,\al} - \Ga^\al_{\ \al\mu,\nu} - K_{\mn}\\
R^{\al}_{\ \nu} &=& g^{\al\mu} R_{\mn}\\ 
R^\ab &=& g^{\be\nu} R^\al_{\ \nu}\\
R &=& g^\mn R_\mn\\
K^\ab &=& g^{\al\mu} g^{\be\nu} K_\mn \\
K &=& g^\mn K_\mn\\
H_\mn &=& R_\mn - \oot g_\mn R\\
H^\al_{\ \nu} &=& g^{\al\mu} H_\mn \\ 
H^\ab &=& g^{\be\nu} H^\al_{\ \nu} 
\end{eqnarray*}
We then have the following relations 
\begin{eqnarray} 
R \sq{g} &=& K\sq{g} + \pa_\mu B^\mu \\
\varep (g_\ab) (K\sq{g}) &\equiv& \left[ {\pa \ov \pa g_\ab} - \pa_\mu {\pa \ov \pa g_{\ab,\mu}} \right] \\
\pa_\mu (\sq{g} H^\mu_{\ \si}) &=& \oot \sq{g} H^\ab g_{\ab,\si} 
\end{eqnarray}
 
\section{Calculus of Variation for a Lorentz-invariant Action} 
 
We collect here the resuts in [3]. 
 
Let $\phi$ represent several multicomponent fields on Minkowski space. 
Summation over Lorentz-, spinor-, and internal indices is always 
implied.  The action is given by 
\begin{displaymath} 
A = \int dx L_0 + \int dx \pa_\mu B^\mu 
\end{displaymath} 
where 
\begin{displaymath} 
L_0 = L_0 (\phi, \pa_\mu \phi), \ \ B^\mu = B^\mu(\phi, \pa_\al \phi), 
\end{displaymath} 
and is assumed to be Lorentz invariant. 
 
The infinitesimal field variation (local variation) is defined by 
\begin{displaymath} 
(\de_* \phi) (x) = \bar \phi (x) - \phi(x). 
\end{displaymath} 
$\de_*$ commutes with the derivative, i.e., 
\begin{displaymath} 
\de_* \pa_\mu = \pa_\mu \de_*. 
\end{displaymath} 
An infinitesimal coordinate transformation on $x^\mu$ results in 
new coordinates $\bar x^\mu$.  The coordinate variation is 
then defined by 
\begin{displaymath} 
\de x^\mu = \bar x^\mu - x^\mu. 
\end{displaymath} 
The variation of a field, induced by a coordinate variation, is given by 
\begin{displaymath} 
(\de \phi) (x) = \bar \phi (\bar x) - \phi(x). 
\end{displaymath} 
We then have the relation 
\begin{displaymath} 
\de_* = \de - (\de x^\mu) \pa_\mu 
\end{displaymath} 
as applied to any field. 
 
The variational principle now reads 
\begin{displaymath} 
\de A = \int dx \, \de_* L_0 + \int dx \pa_\mu 
[ L_0 \de x^\mu + (\pa_\al B^\al) \de x^\mu + \de_* B^\mu ] 
\end{displaymath} 
We now introduce the following abbreviations 
\begin{eqnarray} 
H^\mu &\equiv& \pp {L_0} {\phi_\mu} \\ 
G &\equiv& \varep (\phi) L_0 \equiv \pp {L_0} \phi - \pa_\mu H^\mu 
\quad {\rm Euler\ derivative} 
\\
E^\mu_{\ \si} &\equiv& H^\mu \phi_\si - \de^\mu_{\ \si}{L_0}\quad 
{\rm Energy\ tensor} 
\end{eqnarray}

Then the variational principle becomes 
\begin{eqnarray*} 
\de A &=& \int dx [G \de\phi - G \phi_\si \de x^\si ] \\
&+& \int dx \pa_\mu 
[-E^\mu_{\ \si} \de x^\si + H^\mu \de \phi + B^\mu 
(\pa_\al \de x^\al) - B^\al (\pa_\al \de x^\mu) + \de B^\mu] 
\end{eqnarray*}
For coordinate variations with 
\begin{displaymath} 
\de\phi = - S^\be_{\ \la} (\phi) \pa_\be \de x^\la 
\end{displaymath} 
and the abbreviations 
\begin{eqnarray}
P^{\mu\be}_{\ \ \la} &\equiv& \de^\mu_{\ \la} B^\be 
- \de^\be_{\ \la} B^\mu + \pp {B^\mu} \phi S^\be_{\ \la} 
+  
\pp {B^\mu} {\phi_\be} \phi_\la + \pp {B^\mu} {\phi_\al} 
S^\be_{\ \la} \\
K^{\mu\al}_{\ \ \la} &\equiv& H^\mu S^\al_{\ \la} +P^{\mu\al}_{\ \ \la} 
\end{eqnarray}
The variational principle finally reads 
\begin{eqnarray*}
\de A & =& \int dx 
\left[ - G \phi_\si \de x^\si - G S^\be_{\ \la} \pa_\be \de x^\la \right]\cr 
&+& \int dx \pa_\mu\ 
\left[ -E^\mu_{\ \si} \de x^\si - K^{\mu\be}_{\ \  \la}  
\pa_\be \de x^\la - \pp {B^\mu} {\phi_\al} S^\be_{\ \la} 
\pa_\al \pa_\be \de x^\la \right]\cr 
\end{eqnarray*}
or 
\begin{eqnarray*}
\de  A &=& \int dx \left[ \pa_\be (GS^\be_{\ \la}) -  
G\phi_\la \right] \de x^\la \\ 
&+& \int dx \pa_\mu 
\left[ - \{ E^\mu_{\ \si} + GS^\mu_{\ \si}\}\de x^\la 
- K^{\mu\be}_{\ \ \la} \pa_\be \de x^\la 
- \pp {B^\mu} {\phi_\al} S^\be_{\ \la} 
\pa_\al \pa_\be \de x^\la \right]
\end{eqnarray*}
Translation invariant implies 
\renewcommand{\theequation}{\Roman{equation}}
\setcounter{equation}{0}
\begin{equation} 
\pa_\mu E^\mu_{\ \si} + G \phi_\si = 0 
\end{equation} 
\renewcommand{\theequation}{\arabic{equation}}
\setcounter{equation}{5}
Now raising and lowering indices will be done with the 
Lorentz metric. 
 
We finally introduce the abbreviations 
\begin{eqnarray}
Z^\mu_{\ \si} &\equiv& 
E^\mu_{\ \si} + GS^{\mu}_{\ \si} 
+ \pa_\la K^{\la\mu}_{\ \ \si} 
\\
W^{\la\mu\al} &\equiv&  
\oot 
\left( K^{\mu\al\la} + K^{\al\mu\la} \right) 
- \oot \left( K^{\la\mu\al} + K^{\mu\la\al} \right) 
- \oot \left( K^{\al\la\mu} + K^{\la\al\mu} \right) 
\end{eqnarray}
 
Proper Lorentz invariance together with translation invariance 
now gives 
\renewcommand{\theequation}{\Roman{equation}}
\setcounter{equation}{1}
\begin{equation} 
Z^{\mu\al} = Z^{\al\mu} 
\end{equation} 
\begin{equation} 
\pa_\mu Z^\mu_{\ \si} = - G \phi_\si 
+ \pa_\mu (GS^\mu_{\ \si})+ \pa_\mu \pa_\la K^{\la\mu}_{\ \ \si} 
\end{equation} 
The energy-momentum tensor is given as follows:  Let 
\begin{displaymath} 
t^{\mu\al} \equiv \pa_\la W^{\la\mu\al}. 
\end{displaymath} 
Then 
\begin{displaymath} 
t^{\mu\al} = t^{\al\mu} 
\end{displaymath} 
\noindent 
(i) \quad If \quad $\pa_\mu \pa_\la K^{\la\mu\al} = 0$, then 
\begin{displaymath} 
T^{\mu\al} = Z^{\mu\al} 
\end{displaymath} 
and 
\begin{displaymath} 
\pa_\mu t^{\mu\al} = 0 
\end{displaymath} 
\noindent 
(ii)\quad If \quad $ \pa_\mu \pa_\la K^{\la\mu\al} \neq 0$, then 
\begin{displaymath} 
T^{\mu\al} = Z^{\mu\al} + t^{\mu\al} 
\end{displaymath} 
The energy-momentum tensor is symmetric and is conserved provided 
the equations of motion are satisfied. 

\section{Raising and Lowering Operators} 

\begin{displaymath}
h^{\ab} = \sq{g} g^\ab, h_\ab =  {1 \ov \sq{g}} g_\ab
\end{displaymath} 
 
Many expressions in the theory of gravity are simpler if written 
in terms of the above quantities.  Here raising and 
and lowering is performed by $h^\ab$ and $h_\ab$ respectively. 
 
We introduce the abbreviations 
\begin{eqnarray*}
A^\nu_{\ \mu\la} &\equiv& h_{\mu\be} \pa_\la h^{\be\nu} 
= - h^{\nu\be} \pa_\la h_{\be\mu}\\ 
A_\la &\equiv& A^\mu_{\ \mu\la}
\end{eqnarray*}
Then 
\begin{eqnarray*}
A^\nu_{\ \mu \la} &=& \de^\nu_{\ \mu} \Ga^\si_{\ \si\la} 
- g^{\nu\be} (\Ga_{\be\mu\la} + \Ga_{\mu\be\la})\\ 
A_\la &=& 2\Ga^\si_{\ \si\la} \ \ , \ \  
\pa_\mu A_\nu = \pa_\nu A_\mu\\ 
A^{\nu\si}_{\ \ \la} &=& \pa_\la h^{\nu\si}\\ 
A_{\nu\si\la} &=& - \pa_\la h_{\nu\si}
\end{eqnarray*}
We also have the representation 
\begin{eqnarray*}
\Ga^\mu_{\ \nu\si} 
&=& \oof \{ \de^\mu_{\ \nu} A_\si + \de^\mu_{\ \si} A_\nu 
- h_{\nu\si} A^\mu \}\\
&-& \oot \{ A^\mu_{\ \nu\si} + A^\mu_{\ \si\nu} 
- h^{\mu\al} A_{\nu\si\al} \}\\ 
K \sq{g} 
&=& \ooe A^\al A_\al + \oof  
\left[2 A^{\mu\al}_{\ \ \si} A^\si_{\al\mu} - h^{\al\nu} 
A^{\mu\si}_{\ \ \al} A_{\si\mn}\right] \\
B^\mu &=& - A^{\mu\al}_{\ \ \al} - \oot A^\mu
\end{eqnarray*}
 
\section*{References} 
\begin{enumerate}
 
\item W. Wyss, ``Gauge Invariant Electromagnetic Interactions" 
(to be published).  
 
\item W. Wyss, {\it Helv. Phys. Acta}, {\bf 38}, 469 (1965). 
 
\item W. Wyss, ``The Energy-Momentum Tensor in Classical  
Field Theory" (to be published). 
 
\item M. Carmeli, ``Classical Fields: General Relativity 
and Gauge Theory," John Wiley and Sons, 1982. 

\end{enumerate}

\end{document}